\documentclass[aps,amsmath,amssymb,twocolumn,showpacs,pra]{revtex4}
    \usepackage{graphicx}   \usepackage{bm}
    \usepackage{epsf,times} \usepackage{epsfig}
    \usepackage{dcolumn}    \usepackage{bm,bbm}
    \usepackage[dvips]{color}

\newcommand{\ket}[1]{\mbox{$\left|#1\right\rangle$}}

\newcommand{\bea}{\begin{eqnarray}} \newcommand{\eea}{\end{eqnarray}}
\newcommand{\beq}{\begin{equation}} \newcommand{\eeq}{\end{equation}}

\def\unue#1{{\it#1 ---}}            
                 
\def\openone{\leavevmode\hbox{\small1\kern-4.2pt\normalsize1}}
\begin{document}
\title{Quantum walks on complex networks with connection instabilities and community structure}

\author{Dimitris I. Tsomokos}
\affiliation{Department of Mathematics, Royal Holloway, University of London, Egham, TW20 0EX, United Kingdom}
\date{\today}

\begin{abstract}
A continuous-time quantum walk is investigated on complex networks with the characteristic property of community structure, which is shared by most real-world networks. Motivated by the prospect of viable quantum networks, I focus on the effects of network instabilities in the form of broken links, and examine the response of the quantum walk to such failures. It is shown that the reconfiguration of the quantum walk is determined by the community structure of the network. In this context, quantum walks based on the adjacency and Laplacian matrices of the network are compared, and their responses to link failures is analyzed.
\end{abstract}

\pacs{03.67.Ac, 75.10.Jm, 89.75.Kd} \maketitle

\section{Introduction}
Networks are ubiquitous in both nature and society. They are routinely used to simulate a wealth of phenomena in the physical and biological sciences, as well as in sociology, finance, information and communication technologies \cite{Barabasi_REVIEW,Newman_REVIEW}. In the vast majority of such applications the employed networks are inherently \emph{complex}, by which we mean that there are strong fluctuations in their structural characteristics. This structural disorder is, in fact, a new type of disorder that can lead to cooperative behavior which goes beyond the one encountered in traditional condensed matter physics \cite{Critical_phenomena}.

Quantum networks have become a viable prospect in the area of quantum information processing, with potential applications ranging from teleportation to cryptography \cite{Q_Networks}. In view of their potential use in the foreseeable future, it is clearly beneficial to determine the role of structural complexity in the dynamics of quantum networks.

A small step in this direction is taken in the present work by focusing on a characteristic property of complex networks, which is typically referred to as \emph{community structure} \cite{Community_structure}. Intuitively, a community is a cluster of nodes (vertices) in a complex network (graph), which is connected more densely on the inside than it is connected with the outside. In other words, there are more intracommunity links (edges) within the community than there are intercommunity links between that particular community and other communities in the network.

\begin{figure}[ht]
\includegraphics[width=0.98\columnwidth]{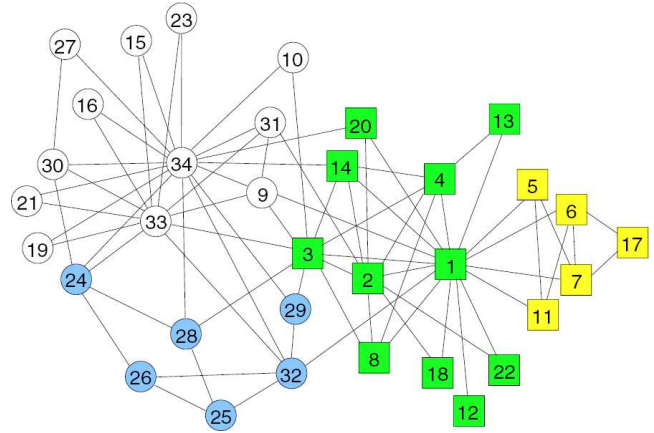}
\caption{(Color online) Community structure in the karate club (KC) network \cite{Zachary} with $N=34$ nodes. The two main communities, centered around nodes $1$ and $34$, are indicated by squares and circles, respectively. Colors correspond to the various possible communities in the network, including sub-communities. Reprinted from Ref. \cite{Fig_paper} (\copyright 2004, IOP Publishing and SISSA).
\label{fig:intro}
}
\end{figure}

As a straightforward illustration we shall examine a social network known as Zachary's \emph{karate club} (KC) \cite{Zachary,Fig_paper}, depicted in Fig. \ref{fig:intro}. The specific network has been studied extensively in the field of community detection \cite{Fortunato_REVIEW}. The main results are presented here in relation to the KC network, but they are valid in general and apply equally well to other networks of increasing size and complexity. In particular, the results have been corroborated by calculations on the \emph{bottlenose dolphins} network with $N=62$ nodes \cite{dolphins} and benchmark \emph{artificial networks} of various sizes $N \in [40, 500]$ with heterogeneous community structure \cite{Andrea}, studied for the altogether different purposes of community detection \cite{Fortunato_REVIEW}.

In this setting, I examine a continuous-time quantum walk (CTQW) and its dynamical response to structural instabilities of complex networks. CTQWs have been studied well in different contexts \cite{Continuous_QRW,CTQW_review}, including quantum search algorithms and quantum communication with spin system dynamics \cite{Quantum_communication}. CTQWs on statistical models, such as small-world networks, have also been studied \cite{CTQW_review}. In the context of spin lattice dynamics \cite{QST_lattices} and modified quantum walks, it was recently shown that quantum walks can detect structural faults in regular graphs \cite{faults,faults_Buzek}. The aim of the present work is different, however, namely to examine the behavior of CTQWs on real-world networks and assess their behavior following a link failure (fault with the connections of the network).

The rest of the paper is organized as follows. Sec. II introduces the model and defines the essential quantities to be used later on. Sec. III solves the model on the KC network and on larger systems, and establishes the main result, which can be quantified using the \emph{node affinity} function, introduced here precisely for this purpose. Sec. IV extends the analysis by comparing the behavior of the system with a different type of quantum walk, in order to assess the robustness in each case (in particular, CTQWs based on the adjacency and Laplacian matrices of a given network are compared and contrasted). Sec. V concludes with a summary of results, comments on experimental implementation, and potential applications in quantum information science.

\section{Model and Definitions}
A complex network ${\cal G}(V,E)$, composed of $N = |V|$ vertices and $K = |E|$ edges, can be described by an adjacency matrix $A({\cal G})$, given by
\bea
A_{ij} &=& 1 \;\;\; {\rm if} \;\;\; (i,j)\in E({\cal G}), \nonumber \\
A_{ij} &=& 0 \;\;\; {\rm otherwise} \nonumber
\eea
for a network that is unweighted, undirected ($A_{ij} = A_{ji}$) and without loops ($A_{jj} = 0$). There are various other matrices that can be associated with a given network \cite{AGT_book}, such as the Laplacian matrix
\bea
L_{ij} = D_{ij} - A_{ij}, \nonumber
\eea
where $D_{ij} = \delta_{i,j} d_i$ is a diagonal matrix determined by the degree of each node $j$, that is,
\bea
d_j \equiv \sum_{i} A_{ij}. \nonumber
\eea

With every node $j = \{1,2,\ldots,N\}$ of the network ${\cal G}$, we associate a basis state $\ket{j}$ in an $N$-dimensional vector space. The basis states are orthonormal and a standard representation can be adopted, such as,
\bea
\ket{1} = \begin{pmatrix}
            1 \\
            0 \\
            \vdots \\
            0 \\
          \end{pmatrix}, \;\;\;
\ket{2} = \begin{pmatrix}
            0 \\
            1 \\
            \vdots \\
            0 \\
          \end{pmatrix}, \;\; \ldots, \;\;
\ket{N} = \begin{pmatrix}
            0 \\
            0 \\
            \vdots \\
            1 \\
          \end{pmatrix}. \nonumber
\eea
Any other state $\ket{\psi}$ can then be written as a linear combination, $\ket{\psi} = \sum_{j} c_j \ket{j}$, where $c_j = \langle j | \psi \rangle$.

At $t=0$ the initial state of the network is $\ket{\Psi(0)}$. At later times, the evolution of the CTQW is given by
\bea
\ket{\Psi(t)} = \exp\left( - i A t \right) \ket{\Psi(0)}.
\label{eq:Psi_t}
\eea
The evolution operator depends on the adjacency matrix $A$ of the network. In the literature, by contrast, it is more common to use the Laplacian matrix instead \cite{Continuous_QRW,CTQW_review}. Therefore, in the penultimate section of this paper, the two CTQWs are compared and contrasted, that is, in addition to the dynamics obtained from Eq. (\ref{eq:Psi_t}) we also examine the evolution according to the equation
\bea
\ket{\Psi(t)}_{\rm L} = \exp\left( i L t \right) \ket{\Psi(0)}.
\label{eq:Psi_t_L}
\eea
Note that the subscript ${\rm L}$ on the state will be used to distinguish it from the state of Eq. (\ref{eq:Psi_t}). In this second case the evolution operator can be decomposed into $\exp\left( i D t \right) \exp\left( - i A t \right)$. Clearly, when the network is regular and each node has the same number of links to other nodes, i.e., $d_j = d$ for every $j$, the first term becomes a multiple of the unit matrix and therefore the CTQWs given by Eqs. (\ref{eq:Psi_t}) and (\ref{eq:Psi_t_L}) are identical, up to an overall phase factor. However, in the case of complex (highly irregular) networks the two evolutions are different.

The probability of finding the quantum walker on a node $j$ at time $t$ is
\bea
P_{j}(t) \equiv |\langle j | \Psi(t) \rangle| ^2
\label{eq:P}
\eea
and we have $\sum_{j}P_{j}(t) = 1$. The time-averaged quantity
\bea
\bar{P}_{j} = \frac{1}{T}\int_{0}^{T} P_{j}(t) {\rm d}t
\label{eq:P_av}
\eea
gives the mean probability of finding the walker on node $j$. In what follows, $\bar{P}_{j}$ plays a crucial role and it is referred to as the \emph{population} of a node $j$.

Finally, unless otherwise stated, we consider that there is an equal \emph{a priori} probability to find the quantum walker on any node $j$ at the start of evolution (at $t=0$). This leads to an initial state of the form
\bea
\ket{\Psi(0)} \equiv \frac{1}{\sqrt{N}} \sum_j \ket{j} = \frac{1}{\sqrt{N}}
        \begin{pmatrix}
            1 \\
            1 \\
            \vdots \\
            1 \\
          \end{pmatrix}.
\label{eq:psi_0}
\eea

In terms of a physical model, the adjacency matrix $A$ is in fact the \emph{effective Hamiltonian} of an $XY$-interaction spin model restricted to the single-excitation subspace. This model has been studied extensively in the field of quantum communication with spin chains and lattices \cite{Quantum_communication,QST_lattices}. Therefore, in this context, the walker is a quantum excitation (i.e., a quasi-particle) diffusing in the network according to Eq. (\ref{eq:Psi_t}) and so, in the following, the term \emph{excitation} is sometimes used to describe the quantum walker.

\section{Main Results}
\subsection{Populations Concentrate on Central Nodes}
One of the most basic structural characteristics of a complex network is the centrality of its nodes. A straightforward measure is the degree centrality,
\bea
C_{j} \equiv \frac{d_j}{N-1}. \nonumber
\eea
For $C_j=0$ the node is isolated; and for $C_j=1$ the node is connected with every other node in the network.

In order to probe the centrality of nodes using the formalism of CTQWs, we begin by assuming that there is equal \emph{a priori} probability $p_0 = 1/N$ of measuring the excitation on any given node. Therefore the initial state of the system is the state $\ket{\Psi(0)}$ of Eq. (\ref{eq:psi_0}). Consequently, the evolved state $\ket{\Psi(t)}$ of Eq. (\ref{eq:Psi_t}) is obtained numerically for a period of time $t\in[0,T]$, in time-steps $\delta t \ll T$, and finally the populations are calculated as prescribed by Eq. (\ref{eq:P_av}).

By comparing the population of each node with its degree centrality we see that populations generically tend to \emph{flow into} highly-connected nodes. This is expected in the sense that, after time $t \gg N$, the average probability of finding the walker on a node $j$ should increase in line with its centrality $C_j$: the more links that are incident on the node, the higher the probability of the walker visiting that node.

\begin{figure}[ht]
\includegraphics[width=0.98\columnwidth]{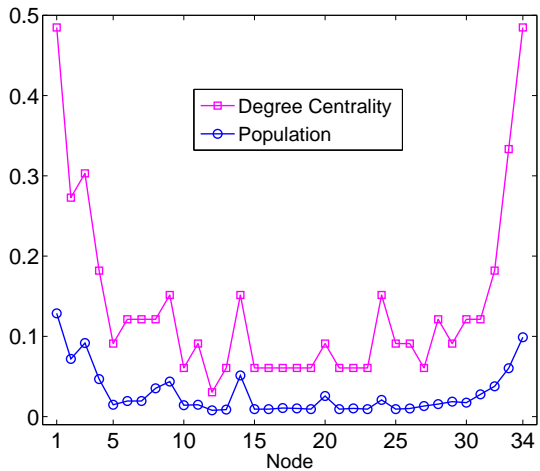}
\caption{
\label{fig:CPvSITE}
(Color online) Degree centrality $C_{j}$ (line of squares) and population ${\bar P}_{j}$ (line of circles) for each node $j = \{1,2,\ldots,34\}$ in the KC network of Fig. \ref{fig:intro}. Parameters used for the numerical simulation: $T=100 \pi$ (dimensionless units) and $\delta t = 10^{-3} T$.
}
\end{figure}

For the KC network in particular, the result is presented in Fig. \ref{fig:CPvSITE}, where both the degree centralities $C_{j}$ and populations ${\bar P}_{j}$ are shown. It is clear that the population distribution is correlated with the degree centrality of the nodes. The result has been corroborated by performing the same numerical calculation on other real \cite{dolphins} and artificial \cite{Andrea} networks. The population outcomes (average probabilities) remain stable with increasing integration times $T$, as long as $T \sim cN$, where $c$ is a positive constant (typically $c \sim 10$).

%

\subsection{Flow of Populations after a Link Failure}
We are now in a position to pose the central question of this work, namely, \emph{If a link fails, how do populations flow?} In other words, the primary focus is on the reconfiguration of average probabilities of the CTQW following the failure (i.e., removal) of an edge from the network.

We saw previously that populations tend to flow out of peripheral nodes and into central nodes. So it is reasonable to expect that population \emph{hubs} are formed around central nodes and that these hubs are sustained by intracommunity edges, connecting members of the community in which the hub belongs. As a result, the removal of an edge strictly from the interior of a community should weaken the hub and therefore the population of the corresponding community as a whole should decrease (while of course the population of the rest of the network should increase). Another way to put the same idea is that the more edges there are inside a community, the more time spent by the quantum walker (e.g., excitation) in that community, and therefore the higher the probability of finding the walker in that region of the network. So if a link fails, this probability is reduced and it becomes more likely to find the quantum walker elsewhere (hence the flow of population out of the community).

\begin{figure}[ht]
\includegraphics[width=0.98\columnwidth]{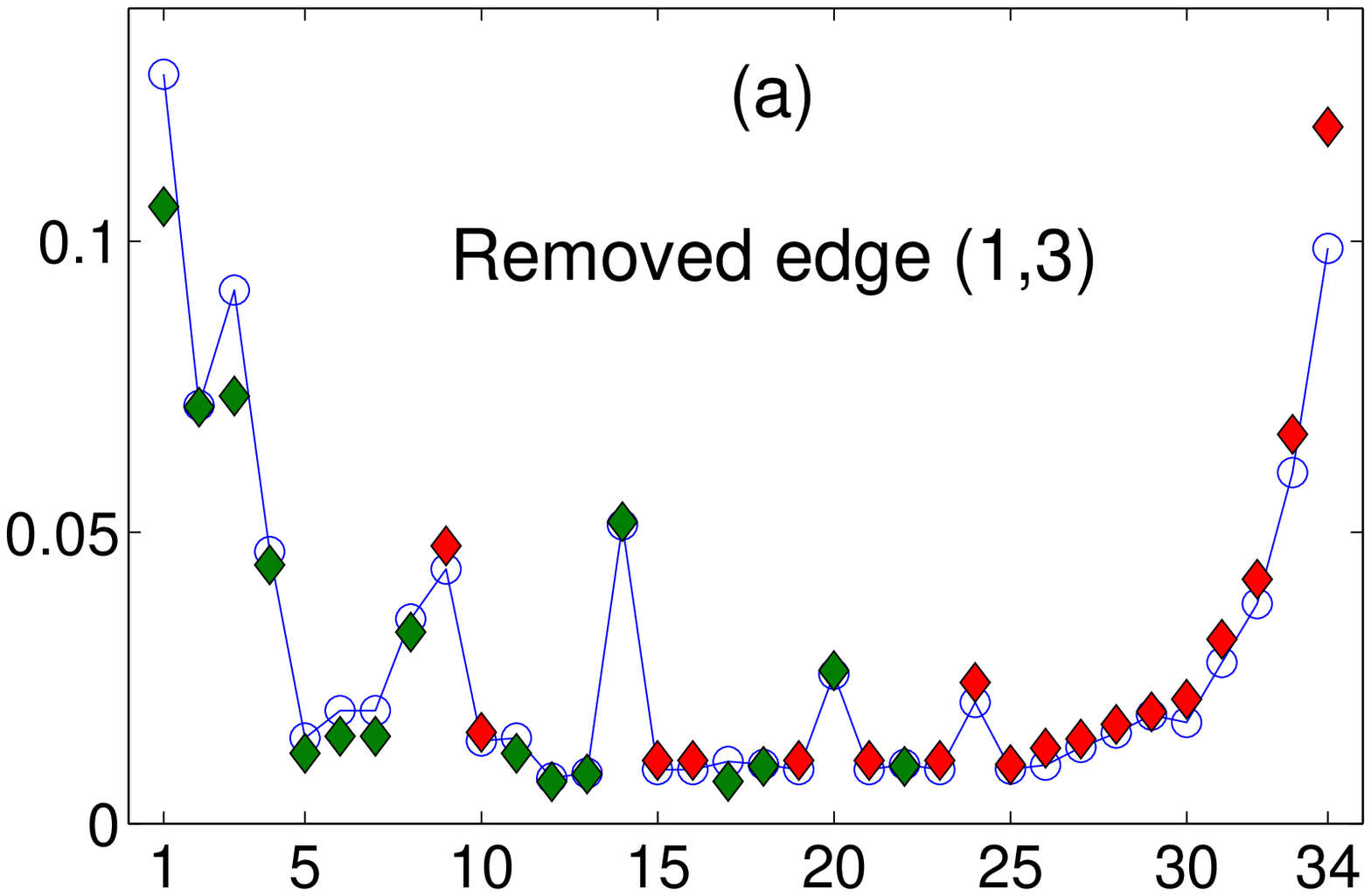}
\includegraphics[width=0.98\columnwidth]{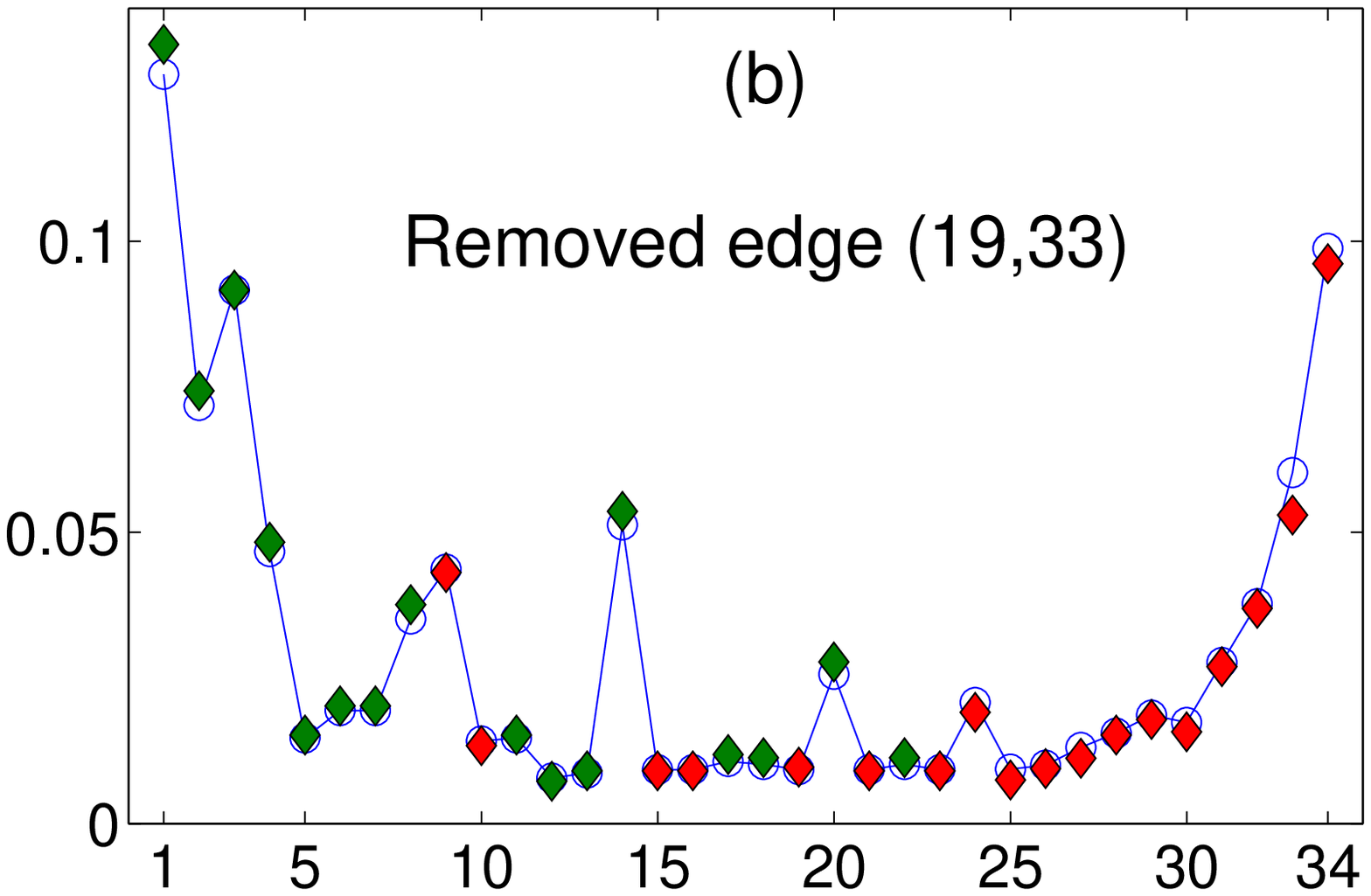}
\includegraphics[width=0.98\columnwidth]{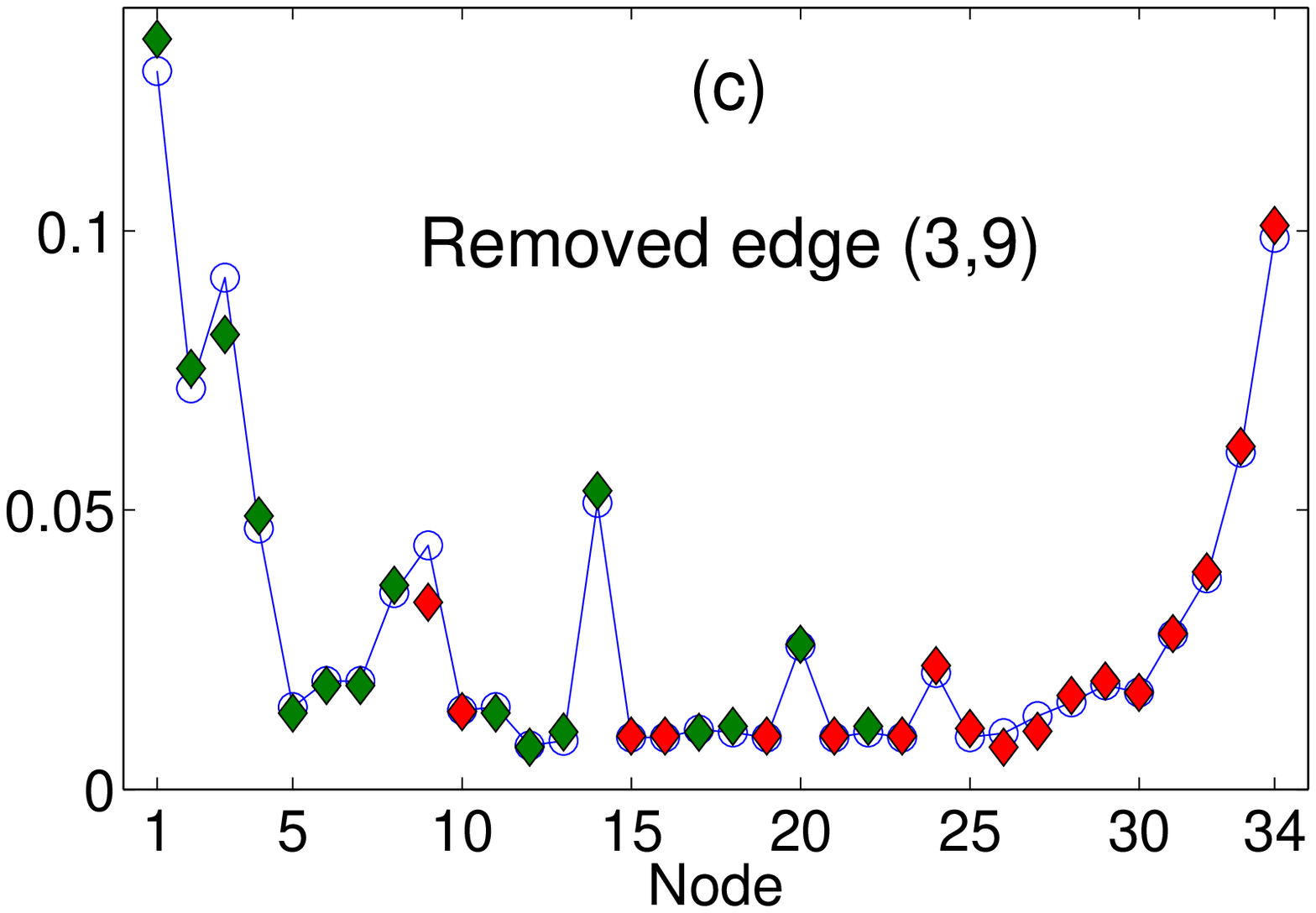}
\caption{
\label{fig:EdgeRemoval}
(Color online) Population flows after an edge is removed. Typical examples are shown: (a) edge $(1,3)$ belongs to the community centered around node $1$ (squares in Fig. \ref{fig:intro}); (b) edge $(19,33)$ belongs to the community centered around $34$ (circles in Fig. \ref{fig:intro}); and (c) edge $(3,9)$ is an intercommunity edge. The (blue) line of circles corresponds to the populations of Fig. \ref{fig:CPvSITE} for an ideal network, while the (green and red) diamonds correspond to the re-calculated (square and circle) populations \emph{after} the link failure.
}
\end{figure}

This expectation has been tested and indeed verified numerically in the KC \cite{Zachary} and other \cite{dolphins,Andrea} networks by removing edges and recalculating populations after the removal operations. Typically, if the edge belongs to a community, populations flow out of that particular community and into neighboring ones. But if the edge removed connects two communities, the populations of both hubs inside these communities are increased, while the population of nodes close to the failed link (removed edge) are reduced.

Three illustrative examples are presented in Fig. \ref{fig:EdgeRemoval}. In Fig. \ref{fig:EdgeRemoval}(a) [\ref{fig:EdgeRemoval}(b)] the failed link belongs to the community indicated by squares (circles) in the KC network of Fig. \ref{fig:intro}; after its removal the populations of the squares (circles) are reduced, as seen by the green (red) diamonds. By contrast, in Fig. \ref{fig:EdgeRemoval}(c) the failed link lies \emph{in-between} the communities of squares and circles; and after its removal the population hubs in both communities are increased.

Therefore, the answer to the main question is that the failure of a link entails that populations flow out of the community in which it belongs and into neighboring ones. If the link is in-between communities, then the populations of both community hubs are increased.

This cooperative behavior, on the community level, has been corroborated with extensive numerical testing on other real \cite{dolphins} and artificial \cite{Andrea} networks. The results are not presented, as they do not add to the main argument, but they are broadly similar irrespectively of the total size of the network. However, in the case of overlapping communities in tailor-made networks \cite{Andrea} the direction of population flow can be more ambiguous, in some special cases, such as those with very high inhomogeneity of the community distribution size. However, such cases are not particularly relevant for the potential applications of our proposal, as these are outlined in the Conclusions.

\subsection{Node Affinity Function}
The previous results mean that the reconfiguration of populations following a link failure entails co-operative behavior on the community level. The populations inside a community flow \emph{in the same} direction when an edge is removed. This motivates us to find a suitable measure of similarity between nodes, conveying the likelihood that two nodes will react in the same way in the event of a link failure.

Therefore, we let the direction of flow on node $j$, after an edge $k \in [1,K]$ has been removed, be given by $\theta_j(k) = \pm 1$, where $+1$ ($-1$) denotes population flowing into (out of) the node. We then define
\bea
\alpha_{ij} = \sum_{k=1}^{K} \frac{\theta_i(k) \theta_j(k)}{K},
\label{eq:affinity}
\eea
which takes values between $-1$ and $1$. Whenever $\alpha_{ij} > 0$ ($\alpha_{ij} < 0$) the two nodes are (are not) likely to react in a similar way, while $\alpha_{ij} = 1$ ($\alpha_{ij} = -1$) implies that they are definitely (not) likely to react similarly. Clearly, $\alpha_{ij}$ is a function capturing the likelihood that two nodes belong to the same community and hence respond to link failures in the same way, so we call it \emph{node affinity}.

\begin{figure}[ht]
\includegraphics[width=0.98\columnwidth]{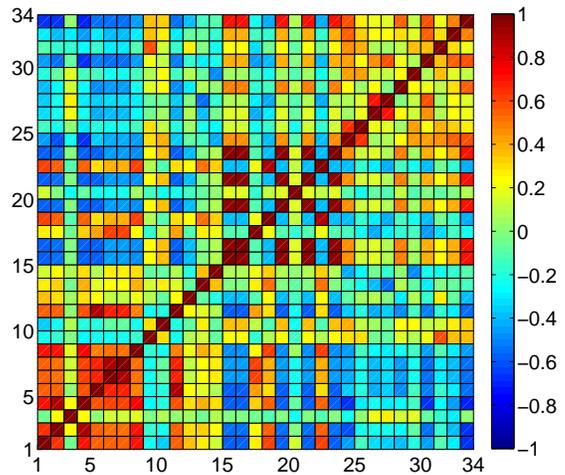}
\caption{
\label{fig:colormap}
(Color online) Node affinity $\alpha_{ij}$ of Eq. (\ref{eq:affinity}) for the KC network of Fig. \ref{fig:intro}. Correspondence between shades (colors) and numerical values is shown in the vertical bar on the right.
}
\end{figure}

The node affinity function for the KC network is presented in Fig. \ref{fig:colormap}. In this case, the function reflects the community structure very accurately indeed. Looking at the first column, for instance, it is seen that node $1$ has high affinity with nodes $2$ to $8$ and $11$ to $14$ (among others), while it has low affinity with nodes $\{ 9,10,15,16,19,21 \}$ and $23$ to $34$. In some other cases, such as large networks with many inhomogeneous communities \cite{Andrea}, the function does not reflect community structure quite as clearly, but it still captures the community-based response to link failures, on average. However, the \emph{similarity} of dynamical response of different nodes to structural instabilities, such as multiple link failures, is captured quite clearly.

%

\section{Comparison with Laplacian-based CTQW}
We now turn our attention to the CTQW determined by the Laplacian matrix of the network, as prescribed by Eq. (\ref{eq:Psi_t_L}). The Laplacian matrix is also known as the connectivity matrix by some authors \cite{CTQW_review} and the corresponding CTQW appears in various areas of physics, chemistry and biology.

The spectrum of the Laplacian matrix can be derived from that of the adjacency matrix only for regular graphs. In the case of complex networks, such as those considered here, the spectra of the two matrices are different. In particular, the eigenvalue spectrum of the Laplacian matrix for a single-component network (i.e., one without isolated regions) is of the form
\bea
l_1 < l_2 \le \cdots \le l_N, \nonumber 
\eea
where the smallest eigenvalue is $l_1 = 0$ since the matrix is positive semi-definite \cite{AGT_book}. Given that $L {\mathbf 1} = 0$, the eigenvector corresponding to $l_1$ is the (normalized) vector ${\mathbf 1}$ with all entries being equal to $1$. This vector is, in fact, the state $\ket{\Psi(0)}$ given by Eq. (\ref{eq:psi_0}); in CTQW notation it is given by ${\mathbf 1} \equiv \sum_j \ket{j}$. Therefore, the initial state $\ket{\Psi(0)}$ will not evolve in time; we have $\ket{\Psi(t)}_{\rm L} = \ket{\Psi(0)}$ for all $t$.

Note that, conversely, if we consider regular graphs and we initiate the CTQW of Eq. (\ref{eq:Psi_t}), which is governed by the adjacency matrix, in the equiprobable state $\ket{\Psi(0)}$ of Eq. (\ref{eq:psi_0}), then the system will not evolve. In this case also, the initial state is an eigenstate of the evolution operator (because the eigenspectrum of the evolution operator based on $A$ is the same as the one based on $L$ for regular graphs).

Consequently, in order to compare the two types of CTQW we need to start from a localized state, $\ket{\Psi(0)} = \ket{j}$. Starting from this state, we calculate the adjacency-type CTQW of Eq. (\ref{eq:Psi_t}) and the Laplacian-type CTQW of Eq. (\ref{eq:Psi_t_L}), and then obtain the long-time average populations from $\ket{\Psi(t)}$ and $\ket{\Psi(t)}_{\rm L}$, respectively, via Eq. (\ref{eq:P_av}).

First, we see that in both cases the populations depend strongly on the initial state $\ket{j}$. The diffusion process is quite different with this initial state, in the sense that much of the initial excitation remains localized in its starting point \cite{CTQW_review}.

Second, in the case of the Laplacian-type CTQW, the populations derived from the long-time average of $\ket{\Psi(t)}_{\rm L}$ do not reflect the (degree) centrality of the nodes in the network. This is due to the strong dependence of the populations on the initial state. To illustrate the point let us assume that we initiate the walk on a weakly-connected node with low centrality; then the long-time average probability of finding the walker on that node will still be high due to the initial conditions, even though the node in question has actually low centrality.

Third, we find that the Laplacian-type CTQW does not respond to link failures (edge removal operations) on the level of community structure. In other words, the deletion of an edge from the network does not cause the populations inside the corresponding community to flow out of that community and into the rest of the network. Instead, we find that populations flow in more complicated ways that do not correlate in general with the community structure of the network.

\section{Conclusions}
We have examined the dynamics of CTQWs on various complex networks \cite{Zachary,dolphins,Andrea}. When the evolution is governed by the adjacency matrix and the walk starts from an equiprobable (delocalized) state, the population of each node reflects its centrality. When a link in the network fails (i.e., if an edge is removed) the populations reconfigure in a way that depends on the community structure of the network.

In particular, if the failed link belongs to a community ${\cal A}$, the populations inside ${\cal A}$ decrease while the populations outside it increase. In other words, populations flow out of the community in which the link failure has taken place. By means of the node affinity function we have quantified the similarity of nodes in their response to such link failures.

By contrast, for CTQWs based on the Laplacian matrix of the network, the dynamics is trivial if we start from the equiprobable state. In this case we need to initialize the walk in a localized state and, as a result, the correlation between node centrality and population is lost. The response to a link failure is not determined by the community structure of the network and so the node affinity function cannot detect similarities of response to link instabilities, among the nodes.

There are various proposals for the experimental implementation of CTQWs on (regular) networks, mostly with quantum optical methods \cite{Continuous_QRW}. Recent proposals for the simulation of exotic lattice systems with superconducting qubits could also be employed, especially for complex network structures, as they allow for arbitrary connectivity between sites \cite{SQ_qubits}.

The results of this work may have applications in quantum network design. The main idea would be that, alongside any other dedicated network function, quantum networks could be monitored in a continuous way by means of the CTQW prescribed by Eq. (\ref{eq:Psi_t}). As long as the network is designed to have straightforward community structure (regular networks included) then the failure of a link, or the failure of neighboring links within a finite area, would be detected by the population flows of the CTQW.

\unue{Acknowledgments}
This work was supported by EPSRC-GB under Grant No. EP/G045771/1. I would like to thank Tobias Osborne for very useful conversations.


\end{document}